\documentclass[journal=apchd5,manuscript=article]{achemso}

\usepackage[version=3]{mhchem} 
\usepackage{xcolor}

\author{Mohammadrahim Kazemzadeh}
\affiliation[]{Center for Biomolecular Nanotechnologies, Istituto Italiano di Tecnologia, via Barsanti 14, Arnesano, 73010, Italy.}
\altaffiliation{These authors contributed equally to this work}

\author{Banghuan Zhang}
\affiliation[]{State Key Laboratory of Precision Spectroscopy, East China Normal
University, Shanghai, 200241, China}
\altaffiliation{These authors contributed equally to this work}

\author{Tao He}
\affiliation[]{State Key Laboratory of Precision Spectroscopy, East China Normal
University, Shanghai, 200241, China}

\author{Haoran Liu}
\affiliation[]{State Key Laboratory of Precision Spectroscopy, East China Normal
University, Shanghai, 200241, China}

\author{Zihe Jiang}
\affiliation[]{State Key Laboratory of Precision Spectroscopy, East China Normal
University, Shanghai, 200241, China}

\author{Zhiwei Hu}
\affiliation[]{State Key Laboratory of Precision Spectroscopy, East China Normal
University, Shanghai, 200241, China}

\author{Xiaohui Dong}
\affiliation[]{State Key Laboratory of Precision Spectroscopy, East China Normal
University, Shanghai, 200241, China}

\author{Chaowei Sun}
\affiliation[]{Institute of Laser Manufacturing, Henan Academy of Sciences,
Zhengzhou, 450046, China}

\author{Wei Jiang}
\affiliation[]{Institute of Laser Manufacturing, Henan Academy of Sciences,
Zhengzhou, 450046, China}

\author{Xiaobo He}
\affiliation[]{Institute of Physics, Henan Academy of Sciences, Zhengzhou, 450046,
China}
\author{Shuyan Li}
\affiliation[]{Tsinghua Shenzhen International Graduate School, Tsinghua University, Shenzhen, 518055, China.}
\alsoaffiliation[]{School of EEECS, Queen’s University Belfast, Belfast, Northern Ireland, BT9 5BN, United Kingdom}

\author{Gonzalo \'Alvarez-P\'erez}
\affiliation[]{Center for Biomolecular Nanotechnologies, Istituto Italiano di Tecnologia, via Barsanti 14, Arnesano, 73010, Italy.}
\author{Ferruccio Pisanello}
\affiliation[]{Center for Biomolecular Nanotechnologies, Istituto Italiano di Tecnologia, via Barsanti 14, Arnesano, 73010, Italy.}

\author{Huatian Hu}
\affiliation[]{Center for Biomolecular Nanotechnologies, Istituto Italiano di Tecnologia, via Barsanti 14, Arnesano, 73010, Italy.}
\altaffiliation{Current address: POLIMA—Center for Polariton-driven Light–Matter Interactions, University of Southern Denmark, Odense M, Denmark}
\email{huhu@mci.sdu.dk}

\author{Wen Chen}
\affiliation[]{State Key Laboratory of Precision Spectroscopy, East China Normal
University, Shanghai, 200241, China}
\email{wchen@lps.ecnu.edu.cn}

\author{Hongxing Xu}
\affiliation[]{State Key Laboratory of Precision Spectroscopy, East China Normal
University, Shanghai, 200241, China}
\alsoaffiliation[]{Institute of Physics, Henan Academy of Sciences, Zhengzhou, 450046,
China}
\alsoaffiliation[]{Institute of Laser Manufacturing, Henan Academy of Sciences,
Zhengzhou, 450046, China}
\email{hxxu@hnas.ac.cn}
\title[An \textsf{achemso} demo]
  {Seeing Beyond RGB Capabilities: {Data-Driven and} Physics-Guided Broadband Spectral Extrapolation of Plasmonic Nanostructures by Deep Learning}

\begin{document}


\abstract{Localized surface plasmons confine light within deep-subwavelength volumes, enabling ultrasensitive near-field responses that underpin a wide spectrum of interdisciplinary technologies. Yet this extreme localization also amplifies unwanted ``noise'' from local nanomorphological variations, resulting in spectral complexities and inconsistencies that have long hindered reproducible and scalable nanophotonics. In this context, optical identification and screening of nanostructures with consistent, target responses offers a practical strategy. However, conventional imaging and spectroscopies, including hyperspectral methods, are limited by a resolution–throughput trade-off, motivating the development of faster, high-precision approaches. Here, we introduce \textit{SPARX}, a deep-learning (DL)-powered paradigm that surpasses conventional imaging and spectroscopic capabilities. \textit{SPARX} batch-{classifies the nanoparticles by their shapes, and }extrapolates broadband dark-field spectra (500--1000 nm) of numerous nanoparticles simultaneously from an information-limited RGB image ($<$700 nm) by learning physical relationships among multiple orders of resonances. Predictions take only milliseconds, achieving a speedup of 2--4 orders of magnitude over traditional methods, while maintaining comparable precision. This transformative, imaging-DL integrated approach enables reproducible nanoplasmonic applications and, more importantly, fundamentally reshapes optical characterization workflows and extends their reach.}

\noindent \textbf{keywords}: {Deep learning, Dark-field, Near-field, Far-field, Single plasmonic nanocavity, Consistency, Predicting spectra from images, Classification}

\maketitle

\section*{Introduction}\label{sec1}
Plasmonic nanostructures, particularly those forming nanogaps, support localized surface plasmons that confine light to deep-subwavelength volumes comparable to molecular and atomic dimensions. \cite{xuSpectroscopySingleHemoglobin1999a,halas2011plasmons,baumbergExtremeNanophotonicsUltrathin2019,benzSinglemoleculeOptomechanicsPicocavities2016a,ciraci2012probing}.
By integrating low-dimensional materials into these nanogaps, they enable probing of electronic (e.g., photoluminescence) \cite{akselrodProbingMechanismsLarge2014a}, vibrational (e.g., surface-enhanced Raman scattering) \cite{langerPresentFutureSurfaceEnhanced2020,benzSinglemoleculeOptomechanicsPicocavities2016a}, and other linear and nonlinear optical properties of  materials \cite{chenIntrinsicLuminescenceBlinking2021, chenContinuouswaveFrequencyUpconversion2021a}, with atom-level sensitivity \cite{giovannini2025electric}, introducing a fascinating field as ``\textit{extreme nanophotonics}'' \cite{baumbergExtremeNanophotonicsUltrathin2019,liBoostingLightMatterInteractions2024a}. 
This high sensitivity arises from the extremely enhanced nanoscopic light-matter interactions supported by nanostructure's quasinormal modes~\cite{li2021bright}, making resonance–material spectral matching a fundamental prerequisite for nanophotonic applications.

However, this extreme localization also amplifies contributions from local nanomorphological features such as nonuniformity, roughness, defects, surface atomic-level dynamics, and pico-cavities inevitably introducing `noise' and compromising reproducibility \cite{wang2023effect,ciraci2020impact,xomalis2020controlling,sun2022revealing,li2021bright,griffiths2021locating,benzSinglemoleculeOptomechanicsPicocavities2016a,bedingfield2023multi,tserkezisHybridizationPlasmonicAntenna2015}.
This intrinsic variability~\cite{hu2023full,baumbergExtremeNanophotonicsUltrathin2019} poses a paradox—enhancing fields for exceptional single-particle performance can inherently undermine spectral consistency and reproducibility across ensembles, constituting a crucial bottleneck in extreme nanophotonics.
One general way of circumventing this is to identify and pre-screen the {target (e.g., spectrally-uniform, with desired resonances)} nanoparticles from countless particles ($\simeq 10^{10}$ per milliliter in the solution) drop-casted on the substrate during the sample preparation, and conduct experiments only on that subset. Conventionally, this screening process has relied on manual expertise combined with imaging and spectroscopic measurements, rendering it inefficient and labor-intensive.

Confocal dark-field (DF) microscopy has long been a cornerstone for single-nanoparticle characterization. Resonances and sample uniformity are often inferred empirically from their images at high throughput, assuming that the color distribution of Airy patterns directly reflects the energy and lineshape of the plasmonic modes. However, human vision cannot reliably resolve subtle chromatic variations, and the information-compressed three-digit RGB output of commercial cameras (400–700 nm) lacks sufficient spectral detail and fails to capture lower-energy resonances. Multiple resonances with distinct far-field patterns further complicate the problem~\cite{pisanello2010room, kongsuwanPlasmonicNanocavityModes2020a}.
More precise techniques, such as DF mapping, acquire spatially resolved spectra via point-scanning, representing a form of hyperspectral imaging (HSI) 
\cite{el-khouryHyperspectralDarkField2016a,zamora-perezHyperspectralEnhancedDarkField2018a,ortegaHyperspectralMultispectralImaging2020,
fakhrullinDarkfieldHyperspectralMicroscopy2021a, mehtaDarkfieldHyperspectralImaging2021, 
bianBroadbandHyperspectralImage2024}. Such sequential acquisition is unsuitable for large-field detection, which can be mitigated with alternative HSI strategies, including line-scanning (push-broom), wavelength-scanning \cite{el-khouryHyperspectralDarkField2016a}, and snapshot techniques \cite{bianBroadbandHyperspectralImage2024}. Recent advances, including compressive sensing \cite{pianCompressiveHyperspectralTimeresolved2017,
xuCompressiveHyperspectralMicroscopy2022,heHighspeedCompressiveWidefield2023,xuCompressiveHyperspectralVideo2024}, have further improved acquisition speed.
However, HSI methods inherently face a resolution–throughput trade-off: higher spectral resolution requires narrower channels, reducing photon flux per channel \cite{wangMultiplexedOpticalImaging2017a, mehtaDarkfieldHyperspectralImaging2021}. This lowers the signal-to-noise ratio and, unless compensated by stronger illumination or more sensitive detectors, demands longer exposure times which is especially critical for weak signals such as single nanoplasmonic structures.
They also entail costly and specialized hardware, large data storage and copious processing requirements, and rigorous calibration protocols \cite{zamora-perezHyperspectralEnhancedDarkField2018a, ahnComparativeStudyAdsorption2019}.
In contrast, confocal DF microscopy combined with RGB imaging remains economical and effective, while readily compatible with emerging techniques (e.g., machine learning, compressive sensing).
However, to fully leverage this approach for high-throughput, precise acquisition of plasmonic nanostructures, the complex and nonlinear correlations between DF images and spectra, particularly beyond the RGB camera’s physical limits, must be revealed.

In this regard, deep learning (DL) has emerged as a transformative tool for decoding complex photonic interactions \cite{wu2025intelligent}. It has demonstrated the ability to learn intricate light–matter correlations through inverse design of plasmonic devices and metasurfaces \cite{montano2025accelerated,kanmaz2023deep, malkiel2018plasmonic, liu2018generative, an2019deep, qiu2019deep, an2022deep, kazemzadeh2025deep}. When integrated with microscopic and nanoscopic techniques, DL has also shown success in noise suppression, signal processing, nanoparticle identification, and super-resolution imaging, which resolves variations and uncertainties at the subwavelength scale \cite{yi2024ai,shiratori2025machine,hu2020single,heNoiseLearningInstruments2024,ju2023identifying,kazemzadeh2024deep,  kazemzadeh2022cascaded,song2022high,lei2024super,lei2024deep}. Overall, DL is particularly valuable for managing uncertainties in experimental and fabrication processes, such as material inconsistencies, structural imperfections, and environmental fluctuations. 

Here, we present a DL framework, named \textit{SPARX} (Spectral Prediction and Reconstruction from RGB with eXtrapolation), that captures the deeper correlations between information-limited DF images and broadband spectral responses. It overcomes the inherent imprecision of human visual perception, the limited information in RGB channels, and the speed constraints of conventional optical characterization. \textit{SPARX} leverages RGB imaging’s maximal photon efficiency (3 broad channels) to capture fast snapshots, and reconstructs the broadband spectra with high throughput (0.4 s/1,000 particles on GPU).
In particular, most of the spectral resonances in our examples occur beyond 800 nm, demonstrating that the \textit{SPARX} model can infer lower-order resonances from higher-order features by understanding their physical relationships, effectively extending spectral predictions beyond the physical capture capabilities of the camera. This can hardly be achieved with simple correlations or empirical experiences, especially with the same high level of precision.

Furthermore, by leveraging heteroscedastic loss, the \textit{SPARX} model could estimate the uncertainty in spectral predictions for individual nanoparticles, effectively quantifying the prediction error. This correlation allows for further refinement in the selection process. {Moreover, as a powerful complementary screening approach, we develop a \textit{SPARX} classifier capable of directly identifying and differentiating nanoparticle shapes from DF RGB images, showcasing an inference capability beyond the reach of conventional spectral acquisition methods.}
All in all, our approach achieves a {$10^{2}$–$10^{4}$-fold} acceleration in characterization speed compared with spectral acquisition, enabling real-time classification of nanostructures in the future.
By replacing spectrometer-dependent workflows with camera-based deep learning, we establish a scalable platform for next-generation optical characterization, nanophotonic device engineering, and biosensor development, where both spectral precision and high-throughput characterization have proven indispensable.

\section*{Results and Discussion}\label{sec2}
\subsection*{Dark-Field Image-Spectra Correlations}
\begin{figure}[!ht]
  \centering
  \includegraphics[width=0.9\textwidth]{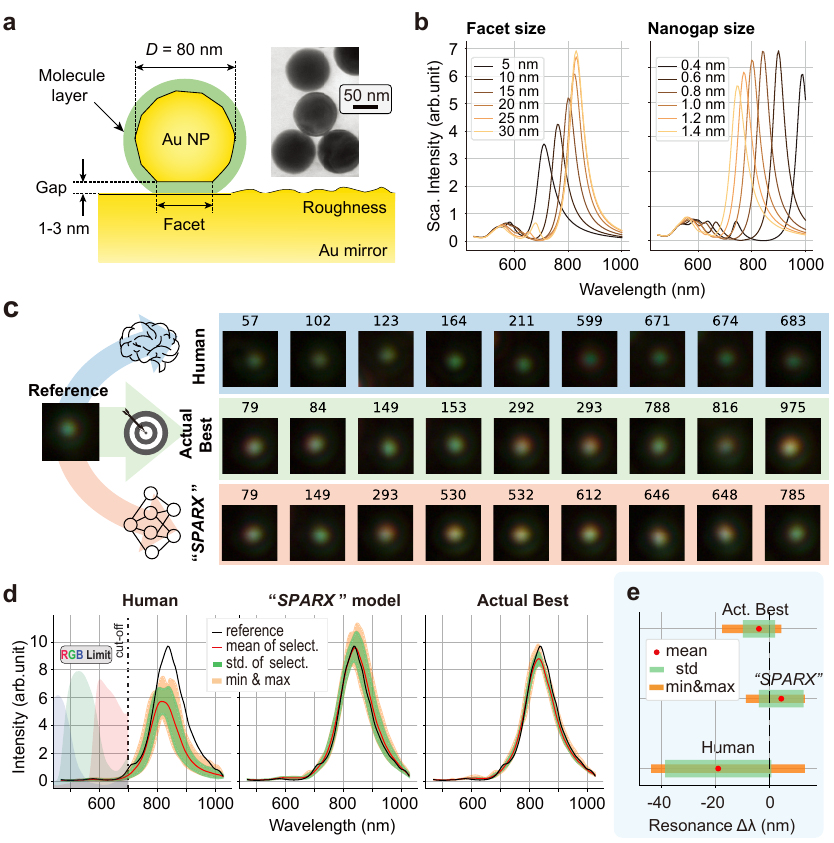}
  \caption{\textbf{Different selecting strategy by the human and deep learning \textit{SPARX} model.} (a) Schematic illustration of the nanoparticle-on-mirror (NPoM) system along with TEM images of the investigated nanoparticles. (b) FEM simulations showing the effect of facet and gap size variations on the resonance of the scattered field. (c) DF images of nanoparticles selected by an expert, our \textit{SPARX} model, and the actual best based on ground-truth spectroscopy. (d) Statistical comparison of spectral deviations between human selection, \textit{SPARX} selection, and actual best from the test dataset. The reference spectrum (black), the mean value (red) and standard deviation (green) of the spectrum selected, min and max values of the selected (orange). The responsive curves of the RGB camera (red, green, blue shades) are overlaid to visualize the detection capacity. (e) Mean, standard deviation, minimum, and maximum of the resonance peak location differences relative to the reference particle, demonstrating that \textit{SPARX} selection closely aligns with the actual best selection.}
  \label{fig:structure}
\end{figure}

Among all plasmonic systems, the nanoparticle-on-mirror (NPoM) configuration (Figure~\ref{fig:structure}(a)) has been widely observed as one of the most versatile platforms, offering easily controllable nanogaps for enhancing optical phenomena \cite{baumbergExtremeNanophotonicsUltrathin2019}. Therefore, we select NPoM as a representative platform for demonstration, measuring 12,000 gold NPoMs to compile a dataset that captures the underlying complexity.
In our experiment, the NPoM system was formed by drop-casting 80 nm cetyltrimethylammonium (CTAC)-capped gold nanoparticles onto a gold mirror (see Methods for sample preparation). The nanogap was defined by a 1-2 nm thick CTAC molecule layer.
From electromagnetic theory, it is well established that the size and shape of the nanoparticles \cite{xomalis2020controlling,bedingfield2023multi}, small variations in the roughness of the metallic surface \cite{ciraci2020impact,wang2023effect}, and the morphology of the gap between the nanoparticles and the metallic substrate (e.g., facets size and adatoms) \cite{tserkezisHybridizationPlasmonicAntenna2015,li2021bright} can drastically alter the resonance behavior of the system.

To visualize these effects, we performed finite element method (FEM) simulations for the DF scattering spectra of the NPoMs using the commercial software COMSOL Multiphysics (see Methods for details). To avoid exhaustive geometric cases discussed above, our simulations focus on two primary geometric factors varying at the nanometer scale: the facet size, and the nanogap thickness between the nanoparticle and the substrate. The results, presented in Figure~\ref{fig:structure}(b), clearly show that even minor modifications to these geometrical parameters lead to significant shifts in the resonances. 
Increasing the facet size may redshift the resonances, which can be related to the polyhedral morphology of the metal's crystalline structure and light-induced atomic migration \cite{xomalis2020controlling}. The nanogap thickness is another crucial factor that may substantially shift the resonances, since the nanogap plasmons have been proven to be able to resolve sub-picometer level of thickness variation \cite{chenProbingSubpicometerVertical2018}. In our test sample, the gap thickness can inherently vary on the level of a few angstroms.

Therefore, due to unavoidable experimental uncertainties, nanoparticles often exhibit variances in spectral responses, especially when coupled with another metallic entity to form a nanogap. Conventionally, this challenge is addressed by selecting optimal nanoparticles based on the DF image appearance, including the colors and shapes of the Airy pattern, by comparing them to a benchmark DF image of a pre-validated "desired" case. In other words, one must first identify the desired nanoparticle based on its spectrum and use its DF image as a reference to look for similar ones.
However, in real-world applications, this empirical screening strategy can often be less accurate. As shown in Fig. \ref{fig:structure}(b) and (d), the strong and primary resonance of interest (around 850 nm) falls beyond the detection capability of the RGB imaging camera. As a result, the DF image features are solely determined by higher-order modes around and below 600 nm, invalidating the conventional reasoning approach.  

To further elaborate on the limitations of empirical human-based selection and highlight the need for DL, we conducted the following experiment. Based on visual perception, we selected the nine DF images that most closely resemble a reference image of a single plasmonic nanoparticle (left panel of Figure \ref{fig:structure}(c)) without knowing the corresponding spectra, under the assumption that spectral similarity follows DF images' visual similarity. The aim was to find the nanoparticle with the spectrum most similar to that of the reference particle. The results, shown in the first row of Figure~\ref{fig:structure}(c), reveal that the nanoparticles selected were all green in their DF images, reflecting an intuitive selection process based on visual similarity. 
Surprisingly, among the 1,000 candidate nanoparticles (test dataset), none of the human-selected particles were among the actual best-matching particles, which were selected based on ground-truth spectra with the minimal mean absolute error (MAE). 
As quantified in Figure~\ref{fig:structure}(d), the actual-best selection has the minimal standard deviations near the reference spectrum (black curve), whereas the human's selection deviates significantly from the reference, showing a blueshift of $\sim$ 20 nm (quantified in Figure~\ref{fig:structure}(e)), a drop in the intensity, and a prominent variation. Strikingly, the hue of DF images of the actual-best selection can also be yellow or orange, highlighting the failure of this empirical-human-based selection rule.


The spectrum of the reference sample (black curve in Figure \ref{fig:structure}(d)) shows a main peak at 820 nm, a shoulder around 710 nm, and weak higher-order resonances around 580 nm. Due to the RGB camera’s cutoff at 700 nm, the most significant spectral contribution comes from the green channel. In contrast, the actual-best selections, with main resonances near 820 nm matching the reference, exhibit orange Airy patterns, as higher-order modes occurring above 600 nm contribute to the red channel (see yellow shades in Fig. \ref{fig:structure}(d) and detailed discussion in Supporting Information (SI) Section S1). These nanoparticles with images in different hues, which, unfortunately, are often overlooked without measuring during empirical selection. This highlights how DF image colors can be misleading, especially with their correlation to resonances beyond RGB limit remaining elusive in complex, imperfect, realistic systems.

The attempt to uncover the correlations among different orders of the plasmonic resonances can be dated back to the Mie theory \cite{bohren2008absorption}, where an analytical solution of the scattering from a spherical nanoparticle can be derived by solving the Maxwell equations. Each Mie resonance (i.e., dipole, quadrupole, etc.) can be intercorrelated through the spherical harmonics. When coupled with its own image through a mirror, the  NPoM plasmonic nanogap can give rise to a richer variety of the hybridized modes \cite{prodan2003hybridization}, with the resonance of $mn$-order mode $\lambda_{mn}$ given by the cylindrical Bessel functions \cite{tserkezisHybridizationPlasmonicAntenna2015}: $\lambda_{m n}=\frac{\pi w n_{\mathrm{eff}}}{J_{m n}-\phi}$, where $J_{mn}$ is $n$-th root of the $m$-th order Bessel function. $n_{\rm eff}$ is the effective refractive index of the metal-insulator-metal junction, and $\phi$ is a proper reflection phase. 
However, this simplified model can only predict the cavity-like resonances of a perfectly spherical nanoparticle with a single ideal circular facet at the bottom gap, with uncertainties such as polyhedral shapes, roughness, nonlocal and quantum effects, and others mentioned above falling outside its scope. Thus, using a simplified, classical analytical solution to extend the spectra of a realistic system beyond the detection limit can be challenging.

In contrast, {although certain dark-field images may appear visually similar, they remain distinguishable in the high-dimensional feature space used by the model. Quantitative comparisons of visually similar image pairs and their corresponding spectra are presented in Fig. S10 in the Supplementary Information.} As given sufficiently representative datasets, DL methods excel at resolving complex and nonlinear correlations with high degrees of freedom. 
In fact, our \textit{SPARX} model has managed to decode the broadband spectrum varying from 500-1000 nm from the images and therefore could select the spectra with the minimal predicted MAE compared with the reference. It hit 3 out of 10 actual best choices as shown in Figure \ref{fig:structure} (c), breaking away from the color-based selection rule. Nonetheless, the rest 7 ``missed'' selections also have a great match with the reference peak, showing a nearly overlapped mean value (Figure \ref{fig:structure} (d)) with low standard deviation.  
Additionally, Figure~\ref{fig:structure}(e) summarizes the mean, standard deviation, minimum, and maximum values of the resonance peak location differences with respect to the reference particle. Again, this analysis shows that the \textit{SPARX}'s selection closely aligns with the actual best selection, exhibiting similar shifts and variances. This reinforces the capability of deep learning in reliably identifying and predicting nanoparticles with optimal spectral properties, outperforming human intuition. Next, we will explain the process of preparing the datasets and training the \textit{SPARX} model.

\subsection*{Acquisition and Unsupervised Analysis of DF Spectra and Images}
\begin{figure}[!p]
  \centering
  \includegraphics[width=0.95\textwidth]{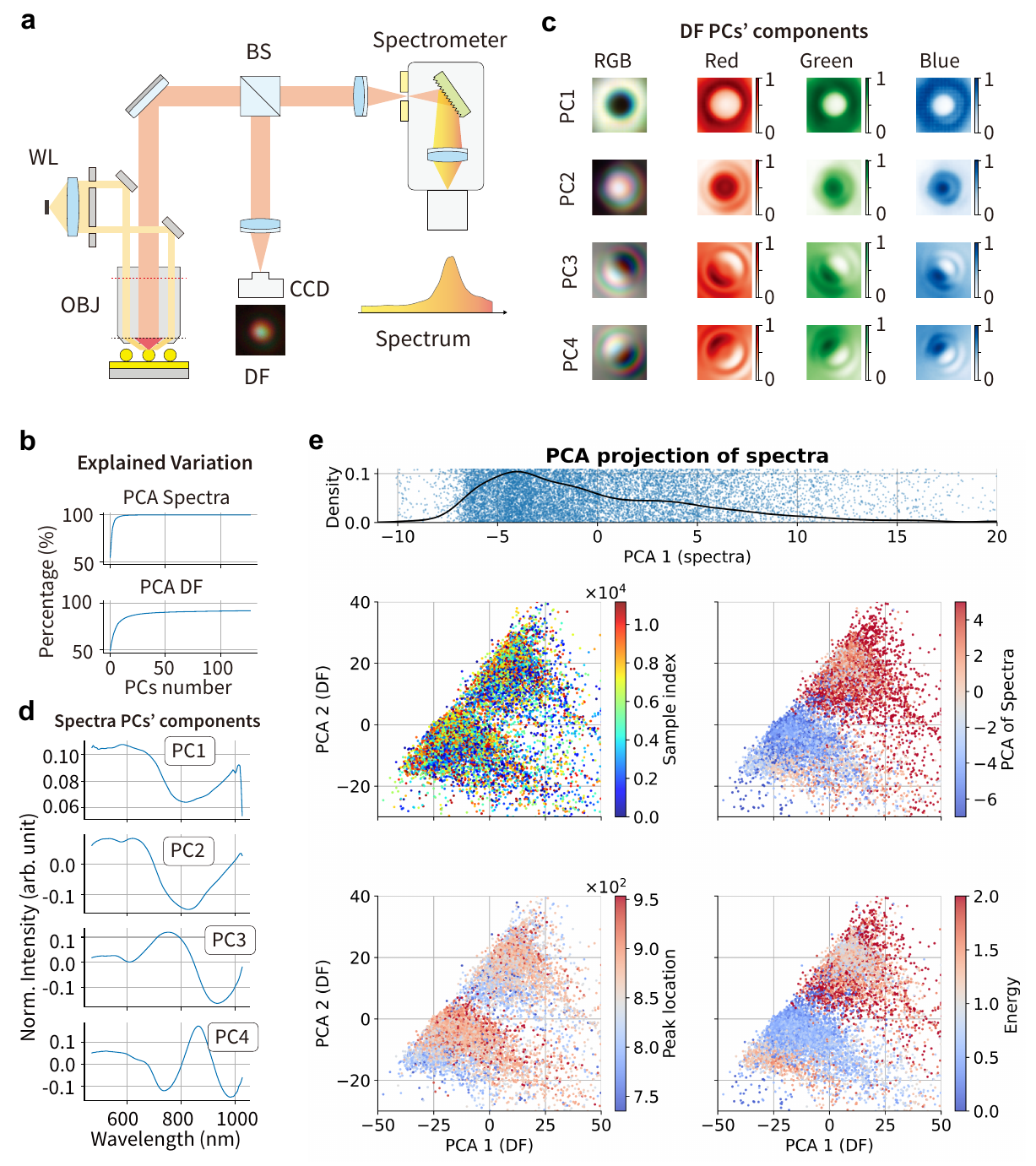}
  \caption{\textbf{Unsupervised learning of the dark-field information.} (a) Optical setup for DF images and spectra collection. WL: white light source. BS: beam-splitter with reflection : transmission in \% (R:T)= 50:50, OBJ: Olympus objective, numerical aperture = 0.9, working distance = 1mm. (b) PCA explained variance, (c, d) First four principal components for DF images and spectra. (e) 1D and 2D PCA projections showing correlations between DF and spectral data, with DF projections colored by the index of the samples (timeline), PCA of spectra, peak location ($\lambda$), and integrated energy.}
  \label{fig:setup}
\end{figure}

The optical setup used to obtain the data for this study is illustrated in Figure \ref{fig:setup}(a) (see details in Methods). It is important to emphasize that an in-house automated measurement protocol was developed to collect the dataset used for training. The dark-field image and the spectrum of each single nanoparticle have been collected simultaneously. More than 12,000 NPoM nanostructures have been measured and captured for the training.

We perform principal component analysis (PCA)  on both the DF images and the processed spectra. PCA reduces high-dimensional data by projecting it onto new orthogonal axes, principal components, that capture most of the variance (i.e., the directions in which the data varies the most or has the most spread). See Methods for data processing and multivariate analysis for more details. The variance explained by the first 128 PCA components for both datasets is shown in Figure \ref{fig:setup}(b). Figures \ref{fig:setup}(c) and (d) display the first four PCA components of the DF images and spectral data, respectively. In Figure \ref{fig:setup} (c), each PCA component is presented with its corresponding red, green, and blue channels. A notable observation is the correlation between the size of extracted features and the diffraction limit of each color channel. For instance, in PC1 of the DF image (Figure \ref{fig:setup}(c), top row), the dark central region in RGB combined plot (where the data is near zero) appears in different colors across the RGB image, but its size decreases from red to blue, consistent with diffraction-limited resolution. Similar trends are observed in other components, such as the concentric ring structures in PC2, which also exhibit size reduction across color channels.
On the other hand, the patterns of the PC1–4 for the DF images clearly present the spherical harmonics (e.g., $l=0$ for PC1, $l=1$ for PC2–4), inferring the NPoM's point scattering nature in the Airy pattern~\cite{pisanello2010room}. These observations demonstrate that our analysis based on PCA captures meaningful physical features, revealing both the optical and structural characteristics of the system. Furthermore, using the PCA of DF images we are able to access scattering information and potentially spectral signatures.

From the spectral PCA components in Figure~\ref{fig:setup}(d), PC1 which captures the largest amount of variance primarily reflects peak shift behavior. It shows higher values when the spectral peak shifts toward the red (1000~nm) or blue (400~nm) regions, with a local minimum around 800~nm.
This behavior aligns with the simulations in Figure~\ref{fig:setup}(b), where minor geometric variations, such as changes in facet size or nanogap thickness, lead to significant spectral shifts. The remaining PCs capture finer spectral details, with alternating positive peaks and negative dips across various wavelengths. Thus, our analysis based on the PCA of DF spectra not only reveals spectral variations but also encodes geometric information.

To explore the correlation between DF images and spectra, we project both datasets into lower-dimensional PCA spaces and visualize the results in Figure \ref{fig:setup}(e). {According to Mie and plasmon hybridization theories, the visible higher-order modes and the near-infrared gap-plasmon modes in NPoM systems originate from the same hybridized eigenmode family, so variations in gap thickness or facet size mainly lead to correlated spectral shifts rather than mode decoupling. This correlation can be evidenced by the data-driven multivariate analysis. }The density distribution of the spectral data along PC1 reveals multiple subpopulations, suggesting that this component captures distinct spectral characteristics. In the 2D PCA projection of DF images, two well-separated clusters emerge (middle panel of Figure \ref{fig:setup} (e)). When coloring the DF image projections using different criteria (see Methods: Multivariate Analysis), clear patterns can emerge.  First, using spectral PC1 projection reveals an alignment between DF image features and spectral variation. Second, using sample indices, reflecting the acquisition timeline, demonstrates the consistency of measurements over time. Third, spectral features, namely the peak location (wavelength $\lambda$) and total spectral energy, further validate that the image-based clustering corresponds to meaningful spectral characteristics. These patterns highlight the physical relevance of the clusters found in our analysis (Figure~\ref{fig:setup}(e))

To further support these observations, we applied Uniform Manifold Approximation and Projection (UMAP), a nonlinear dimensionality reduction technique, to both spectral and DF image data, as shown in {SI Figure~S2}. The spectral data is embedded into a one-dimensional UMAP space, while the DF images are projected into two dimensions. By coloring the DF UMAP projection using various spectral-derived features including UMAP values of the spectra, peak positions, and total spectral energy, we uncover strong visual correlations between the DF and spectral modalities (2D maps in {SI Figure~S2}). An interesting outcome is the appearance of multiple distinct clusters in the DF UMAP space (in contrast to PCA projection in Figure~\ref{fig:setup} (e)), which exhibit clear correspondence to specific spectral features. For example, the cluster located at the bottom right of the projection space shows a color gradient where resonance wavelength and total spectral energy vary in opposite directions, i.e. as the resonance red-shifts, the intensity decreases. Additionally, when coloring the projections based on sample index, correlated with the data acquisition time, we observe no discernible pattern, confirming the high reproducibility and stability of the measurements across multiple days. This improved clustering allows us to reveal the rich spectral information that DF images inherently carry, which is typically inaccessible in traditional DF imaging analysis.

The application of unsupervised data analysis techniques reveals a strong correlation between DF images and the spectra, suggesting that a meaningful connection should help with the spectra prediction. This success motivates us to apply supervised models to further refine and strengthen this relationship.

\subsection*{Supervised Deep Learning with \textit{SPARX}: Modeling the Heteroscedasticity and Outliers}

To apply our DL model, \textit{SPARX}, to this dataset, we need to design an architecture that translates 2D DF RGB images into 1D spectral data (see SI, Section S3). This autoencoder-based architecture includes convolutional encoder-decoder components and residual connections to preserve spatial and spectral features. 

We evaluate the model using MAE loss, which demonstrates the network’s ability to capture key spectral features, such as lineshape and intensity (see SI Fig. S{4}). We analyze model performance based on training data size and compare reconstructed spectra with the ground truth (see detailed error distributions and performance benchmarks in SI, Section S{4}). 
However, we find that the performance varies across the spectral range. Specifically, the error varies across different wavelengths. This variability is likely due to limitations in the DF images and the complex, wavelength-dependent physics of nanoscale systems. Such violation of the assumption of constant variance in prediction errors is called heteroscedasticity. 


To address the wavelength-dependent uncertainty, we implement a heteroscedastic learning strategy based on the same architecture (SI Fig. S{3}) through a probabilistic reformulation of the loss function. Instead of relying on a fixed metric like MAE, we model the prediction error at each wavelength as a Gaussian distribution. The model is thus designed to output both the \textit{mean} and the \textit{uncertainty} (variance) of the prediction for each wavelength, that is, how certain the model is about its prediction. During training, the model minimizes the negative log-likelihood (NLL) of this distribution, enabling it to account for wavelength-dependent errors and estimate prediction uncertainty on a per-sample basis. This approach not only improves model robustness but also provides a statistical interpretation at the single-nanoparticle level. The full mathematical formulation is detailed in the Methods section.

\begin{figure}[!ht]
  \centering
  \includegraphics[width=\textwidth]{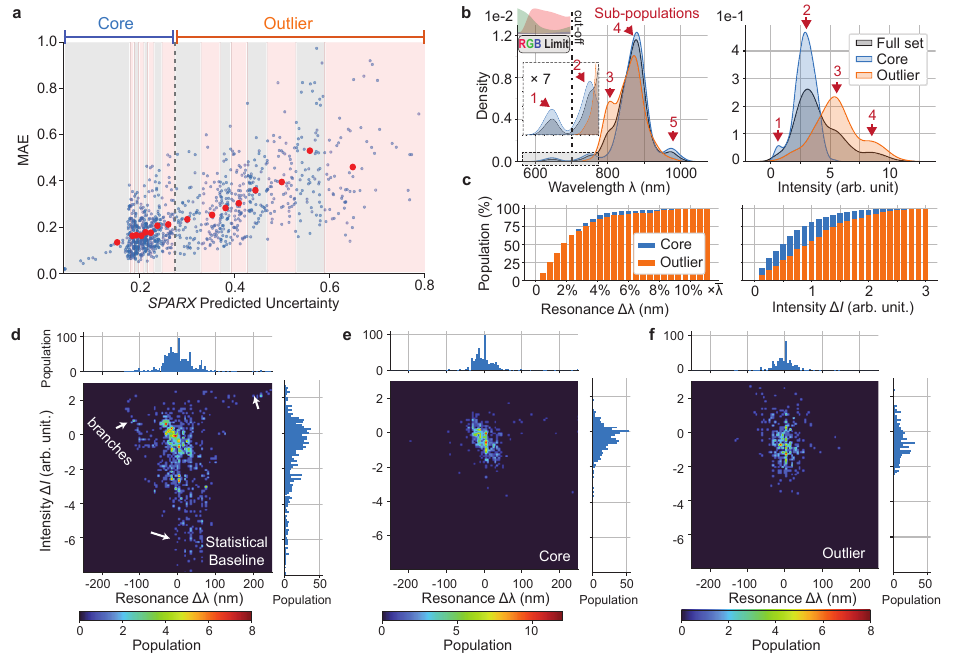}
  \caption{\textbf{Uncertainty-based performance analysis of the \textit{SPARX}.} (a) Scatter plot of prediction error vs. model uncertainty with binned subgroups highlighting the correlation. We set the \textit{SPARX} predicted uncertainty, corresponding to an averaged MAE $\simeq 0.2$, as the threshold for dividing the core and outlier datasets. Red dots are the center of mass of each bin. DL uncertainty has a near-linear correlation with the MAE. (b) Distribution of resonance wavelength and intensity for core (low-uncertainty) and outlier (high-uncertainty) datasets. Inset of (b) shows the zoomed-in (7 times) part of small features near the short-wavelength region surrounded by a dashed box. The dash-dot line indicates the RGB detection capability below 700 nm, with the camera's RGB responsive curve overlaid above. Interestingly, peaks with locations below 700 nm are all categorized into the core set as confident data by \textit{SPARX}. Subpopulations are marked by red arrows. (c) Cumulative histograms of prediction errors for resonance and intensity, showing sharper error convergence in the core group. (d–f) Joint error distributions for baseline model, and the \textit{SPARX} performance over core, and outlier datasets.}
  \label{fig:6-heterosce-loss}
\end{figure}

Figure \ref{fig:6-heterosce-loss}(a) presents a scatter plot comparing the prediction error (i.e., MAE) to the model’s estimated uncertainty for each nanoparticle in the test set (blue points). Model uncertainty is defined as the mean predicted standard deviation across all wavelengths. Notably, a positive correlation is observed between prediction error and uncertainty, suggesting that the model is capable of estimating its confidence.

To further highlight this relationship, we divide the test set into 16 bins based on ascending uncertainty levels. Each bin is visualized as a colored strip in the scatter plot, and the center of mass of each bin is marked with a red dot. These centers follow a near-linear trend, indicating that prediction error increases with uncertainty. This insight enables the identification of a subpopulation of nanoparticles for which the model is highly confident in its predictions. By selecting the first eight bins with the lowest predicted uncertainties, we can isolate predictions with significantly lower average MAE (i.e.,  average $\mathrm{MAE}\leq0.2$), facilitating more reliable spectral reconstructions.

Figure \ref{fig:6-heterosce-loss}(b) presents the distributions of resonance wavelength and intensity for both the core dataset (the first eight bins) and the outlier dataset (the remaining bins). Five distinct sub-population clusters are observed in the wavelength domain (left panel), likely corresponding to the five most probable geometric configurations influencing scattering behavior. An important observation therein is that the distribution of resonance wavelengths within the core dataset closely follows the overall distribution (full set), suggesting that \textit{SPARX}'s selection based on uncertainty does not restrict the diversity of spectral features. 
Strikingly, all data points with resonances occurring in the RGB camera's capacity (below 700 nm) are exclusively selected as part of the core dataset (see the inset of the left panel). This aligns with one's empirical expectations, as RGB-based DF imaging inherently captures more information about resonances within the camera’s detection range, allowing the model to make more confident predictions. In contrast, spectral information beyond the RGB limit depends on inference and extrapolation, which naturally introduces uncertainties. 

In contrast, the intensity distribution, shown in the right panel of Figure \ref{fig:6-heterosce-loss}(b), presents a distinctly different behavior. The overall population density plot of resonance intensity (shown in black) displays three distinct peaks, indicating the existence of three subpopulations. Interestingly, the core group is primarily composed of the subpopulation with the lowest resonance intensities. The remaining two subpopulations with higher intensities are predominantly classified as outliers. Interestingly, the spectra with modest intensity ($\simeq$2.6 arb. units), which contain the largest number of samples, are most recognized as the core data that \textit{SPARX} is most confident in. This may be attributed to their generally more uniform and well-behaved geometries, with reduced randomness stemming from previously discussed factors. 
In contrast, the higher-intensity subpopulations are treated as outliers by \textit{SPARX}, likely due to more complex features such as sharp polyhedral facets, nonuniform gaps, or even picocavities. 
These features have more randomness in general, and are more difficult to characterize using only DF images, leading to increased uncertainty in their prediction.  {That means, the identification of outliers is primarily data-driven rather than based on direct structural evidence from SEM or TEM, since the features inside the nanogap are typically inaccessible by SEM or TEM.} 
In fact, this also aligns with the empirical intuition in plasmonics and surface-enhanced Raman scattering: extreme enhancement with high intensity often compromises robustness and consistency. Nanoparticles exhibiting such exceptionally high intensities remain rare outliers relative to the overall population. Remarkably, our \textit{SPARX} model captures this trade-off by analyzing the associated uncertainty.
It is important to note that while our heteroscedastic model is capable of identifying such complex cases from DF images, accurately reconstructing their spectral details remains challenging due to the limitations of the input modality. Detection of these outliers should not be conflated with precise spectral prediction.

To further demonstrate how the model's certainty translates into improved predictions, we analyze the distribution of prediction errors for both resonance wavelength and intensity. This is visualized in Figure \ref{fig:6-heterosce-loss}(c), where the x-axis represents the error magnitude and the y-axis shows the percentage of the dataset with errors below that threshold. For instance, in predicting the resonance peak wavelength, over 90\% of the core data exhibits a resonance prediction error of less than 4\% of the mean resonance ($\bar{\lambda}$). Additionally, the cumulative histograms reveal that the core dataset shows a steeper increase in percentage at lower error values for both resonance and intensity. This clearly indicates that the model provides significantly more accurate predictions for the core subset compared to the outlier group, further reinforcing the utility of uncertainty-based screening in practical applications.

To explore the link between model uncertainty and prediction accuracy, in Figures \ref{fig:6-heterosce-loss}(d-f), we generated joint plots of resonance wavelength error versus intensity error. Also see SI, Fig.~S{5} for the comparison of the joint plots for the homoscedastic \textit{SPARX} model. As shown in Figure~\ref{fig:6-heterosce-loss}(d), we used the dataset’s mean spectrum as a statistical baseline—a naive yet reasonable choice in the absence of a predictive model. Notably, this represents a statistical reference and should not be confused with a true "prediction," as it cannot reconstruct spectra from individual images. Consequently, its distribution can be primarily determined by the uniformity of the NPoM samples. This baseline exhibits large errors in both wavelength and intensity, with distinct branching in the error distribution, suggesting a nonlinear correlation between resonance wavelength and intensity. 
These branches are notably absent in the prediction errors of \textit{SPARX}, {shown in Figures~\ref{fig:6-heterosce-loss}(e) and (f)} for the core and outlier datasets, respectively. This suggests that the model has learned and leveraged these underlying relationships for improved spectral prediction. Furthermore, the core dataset displays a much tighter error distribution, confirming that predictions are more accurate for data points associated with lower model uncertainty.

\begin{figure}[!ht]
  \centering
  \includegraphics[width=\textwidth]{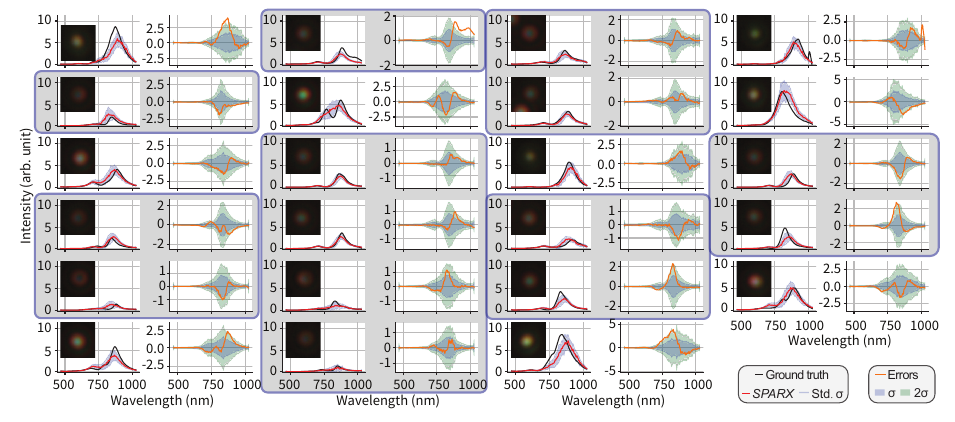}
  \caption{\textbf{\textit{SPARX} spectral reconstruction and outlier prediction.} Example spectra with predicted mean (red), ground truth (black), and uncertainty intervals (blue bands), demonstrating the model’s ability to quantify prediction confidence. All the examples in the purple boxes are data points categorized by \textit{SPARX} as part of the core dataset.}
  \label{fig:6-heterosce-lossb}
\end{figure}

In addition, Figure \ref{fig:6-heterosce-lossb} showcases 23 randomly selected examples (which are the same ones predicted by the homoscedastic model as shown in the SI, Fig. S{4} from the test dataset), each accompanied by its corresponding DF image, \textit{SPARX} heteroscedastic-predicted spectrum, and ground truth spectrum. The $y$-axis (intensity) of each DF spectrum is fixed to the same range, illustrating that \textit{SPARX} can faithfully reconstruct both spectral intensity and wavelength with minimal errors.
The predicted uncertainty is shown as blue shaded bands around the mean prediction, representing the standard deviation. Notably, when the \textit{predicted uncertainty} is considered, the ground-truth spectra largely fall within this range. It means that, for the same DF images, the heteroscedastic \textit{SPARX} model can resolve single-particle-level uncertainties.

To better illustrate the relationship between prediction error and uncertainty, each DF spectrum is accompanied by a panel (on the right) showing the absolute error (orange) between the ground truth and predicted mean. The predicted uncertainty bounds, corresponding to one and two standard deviations ($\sigma$, $2\sigma$), are overlaid for comparison.
These confidence intervals approximate the 65\% and 95\% confidence levels, respectively, providing a clear, visual indication of how well the model's predicted uncertainty (blue and green shades) captures the actual reconstruction error.
In real-world applications, one can use \textit{SPARX} to screen out the most similar spectra with the least predicted uncertainties. This will significantly increase the screening efficiency.

{Beyond NPoM, we further validated the generality of \textit{SPARX} on a more complex plasmonic system that is nanopatch antennas. This system is composed of single $\sim$75~nm gold nanocubes on a mirror  (i.e., NCoM, see SI Section S5, Figs. S6, S7).
Unlike nanospheres, nanocubes produce more diverse spectral responses, as geometric variations can generate multiple resonances associated with different gap sizes and bottom facet geometries \cite{lassiter_plasmonic_2013,pellarin_fano_2016,chikkaraddy2017ultranarrow}. Their synthesis also tends to yield more complex and less uniform shapes, adding to the spectral diversity. Despite the intrinsically higher variability of NCoM spectra (SI Fig. S6(b)) compared to NPoMs (Fig. \ref{fig:6-heterosce-loss}(d)), \textit{SPARX} consistently achieved substantially lower prediction errors than a naive statistical baseline (SI Fig. S6(c)), demonstrating that \textit{SPARX} serves as a general framework capable of handling diverse plasmonic nanostructures beyond spherical geometries.}

\begin{figure}[ht]
  \centering
  \includegraphics[width=0.75\textwidth]{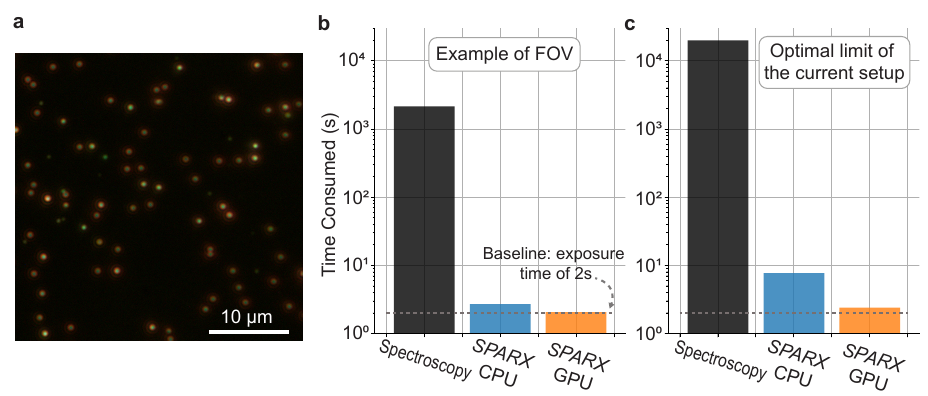}
  \caption{\textbf{High-throughput prediction by the \textit{SPARX} model.} (a) A full scope of the Airy patterns of each nanoparticle in the dark field of view (FOV). Each Airy pattern can be used to predict the corresponding scattering spectrum. (b) and (c) Comparison of the time consumed by different methods: measuring each nanoparticle by spectroscopy (black), predicting with the \textit{SPARX} model run on CPU (blue) or GPU (orange). A case with specific FOV (b) and with the theoretical optimal limit which considers the most compact non-overlapped collection (c). The time baseline 2s indicates the exposure time of the RGB camera to take a photo as (a).}
  \label{fig:7}
\end{figure}

So far, we have analyzed the reliability of \textit{SPARX} predictions for spectral reconstruction. Now, we quantitatively compare its efficiency with conventional spectrum acquisition methods in Figure \ref{fig:7}. Our fully automated spectroscopy system requires an average of 25.1 seconds per nanoparticle, including stage movement, focusing, and spectral acquisition.
However, a single field of view (FOV = $(34.67\,\mu \textrm{m})^2$ in this case) in DF microscopy can contain 10\textsuperscript{1}–10\textsuperscript{3} nanoparticles, depending on the nanoparticle density. In the example shown in Figure \ref{fig:7}(a), 68 scattered nanoparticles with distinct Airy patterns were recorded. The comparison of the time required for individual spectroscopy with \textit{SPARX}-based predictions in Figure \ref{fig:7}(b) highlights a substantial speed advantage. Running our \textit{SPARX} model on an NVIDIA 3090 Ti GPU or an Intel Core i7 12700 CPU yields a speed-up of three orders of magnitude over traditional spectroscopy.

The theoretical upper limit of nanoparticles processable per FOV in our setup is estimated based on the non-overlapping spectral collection area of $\simeq$ 1.2 $\mu$m\textsuperscript{2} per nanoparticle (3–4 pixels of the Horiba spectrometer, with one pixel calibrated as $\simeq$ 0.3 $\mu$m). Under optimal conditions in Figure \ref{fig:7}(c), $\sim$1000 nanoparticles can be processed in a single DL prediction cycle. Processing $\sim$1000 DF images with GPU takes only ~0.4 s, suggesting a potential four-order magnitude speed improvement compared to conventional {confocal DF spectroscopy. Furthermore, while state-of-the-art HSI techniques can already achieve a one–two order of magnitude improvement over point-scanned confocal methods, our \textit{SPARX}-empowered RGB imaging still provides an overall acceleration exceeding two orders of magnitude, depending on the reference baseline.}

A key bottleneck in Figure \ref{fig:7} is the exposure time of our RGB camera (2s per snapshot {for high resolution}), as \textit{SPARX}'s inference itself takes only fractions of a second (Figure \ref{fig:7}(b)). {Future improvements in throughput and performance can be approached from both algorithmic and optical perspectives. Algorithmically, \textit{SPARX} inference is already fast, but accuracy and uncertainty estimation could be enhanced by refining the heteroscedastic loss function, exploring alternative error distributions (e.g., heavy-tailed or skewed), optimizing network architectures and hyperparameters, or adopting advanced models such as variational autoencoders (VAEs) or Bayesian neural networks (BNNs) to better capture stochastic single-particle spectral responses.}
{Optically, acquisition speed could be improved by increasing camera gain (leveraging the model’s noise robustness), expanding the field of view with lower-magnification objectives or wide-field illumination, boosting illumination intensity using high-brightness sources, and enhancing spatial resolution with higher NA objectives, confocal configurations, or advanced techniques such as structured illumination microscopy \cite{heintzmannSuperResolutionStructuredIllumination2017}, super-resolution plasmonics \cite{willetsSuperResolutionImagingPlasmonics2017}, or DF nanoscopy \cite{lee2022hyperbolic}. These combined algorithmic and optical strategies can significantly accelerate measurements while maintaining or improving spectral prediction accuracy.}

{From the perspective of information acquisition,} spectral optimization presents two complementary avenues. Firstly, one can tailor the plasmonic systems to exhibit dominant resonant peaks within the visible range (400–700 nm). It would align spectral signatures with the RGB camera’s detection window, maximizing signal capture efficiency and allowing significantly shorter exposure times. Secondly, inspired by hyperspectral techniques \cite{zhang2022survey}, a few detection channels beyond 700~nm can be incorporated—for example, via multi-channel split-frequency systems such as grayscale CCDs with spectral filters—to broaden the spectral window up to 1000~nm or beyond, thereby capturing near-infrared resonances.

\subsection*{Beyond Spectral Prediction: Shape Classification from DF RGB Images}
{Beyond spectral reconstruction, DF RGB images encode additional information that \textit{SPARX} can exploit for nanoparticle shape classification.  Since metallic nanoparticle's shape strongly influences plasmonic performance \cite{noguez2007surface}, and synthesis often yields by-products with varied geometries, this approach provides a valuable, noninvasive alternative to conventional techniques such as scanning electron microscopy (SEM), transmission electron microscopy (TEM), and atomic force microscopy (AFM). In future applications, \textit{SPARX classifier} could distinguish multiple particle geometries simultaneously on the same substrate, enabling additional rapid screening of desired nanoparticles or studies that leverage spectral diversity from different particle types \cite{gschneidtner2020constructing}.}

{The underlying physical motivation for classification lies in the pronounced differences in the spatial distributions of near- and far-field among nanoparticles of different shapes (e.g., nanocubes and nanospheres shown here). These differences are expected to be reflected in their DF RGB images (SI Fig. S8(a)). Again, these RGB images can be challenging for human vision to distinguish, as visually similar images may belong to different classes. Before developing a supervised classification model, we first conducted an unsupervised study to explore the natural separability of the two datasets (SI Fig. S8.  
We normalized each RGB image to its maximum pixel value, removing the influence of acquisition time and variations in optical collection efficiency. This ensures that any clustering arises from intrinsic image features rather than overall brightness. We also plotted the intensity distributions for each RGB channel in Fig.~S8(b), which display similar patterns. 
The UMAP distribution RGB images in Fig.~S8(c) shows that the nanospheres are relatively separable from nanocubes. These findings support the feasibility of using a supervised model to achieve robust nanoparticle classification.}

\begin{figure}[ht]
  \centering
  \includegraphics[width=\textwidth]{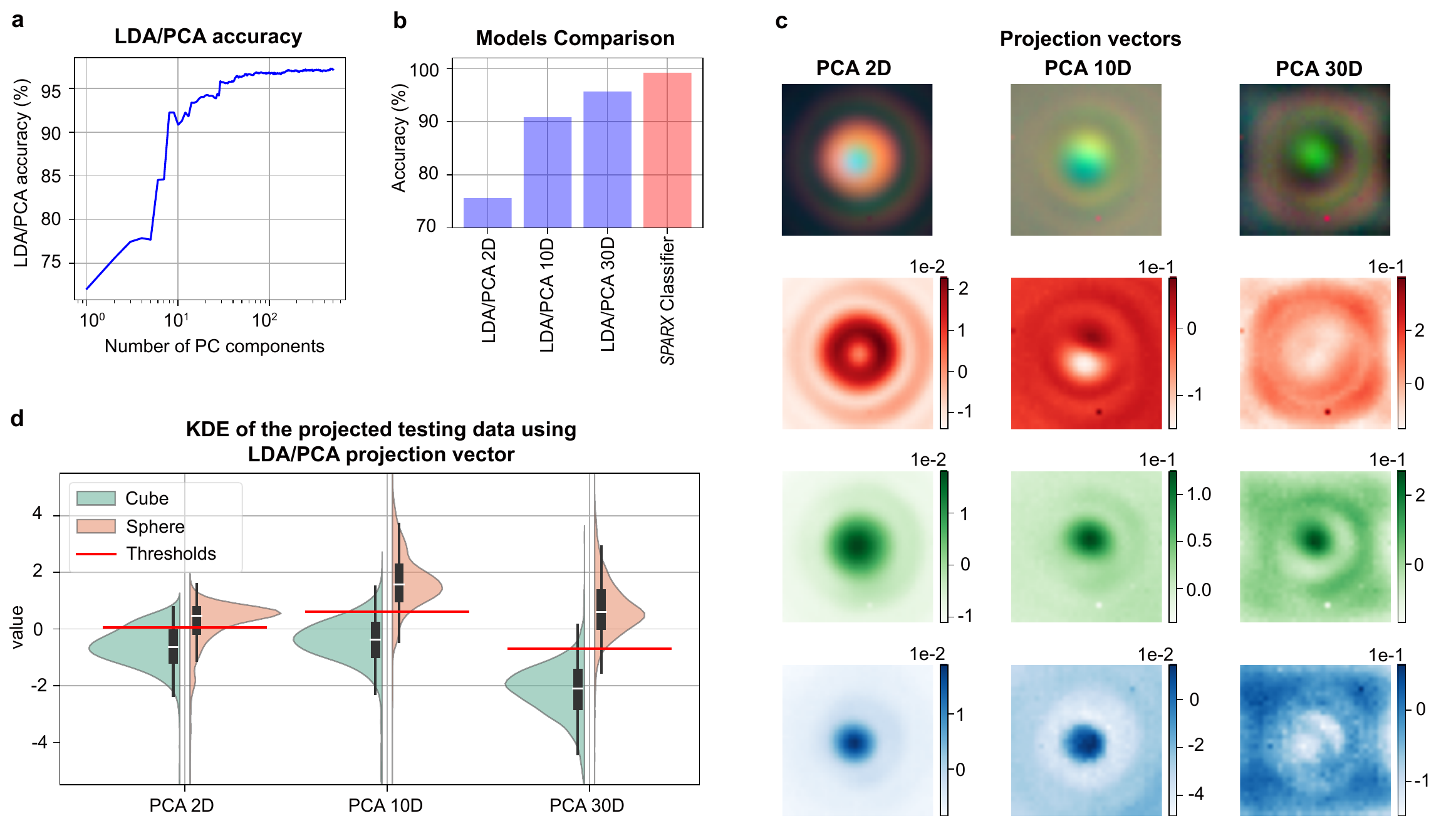}
    \caption{\textbf{PCA–LDA classification of nanospheres and nanocubes from DF RGB images.} 
    (a) Classification accuracy as a function of the number of PCA components used before LDA, showing significant performance gains at 10 and 30 components. (b) The comparison of the accuracy of different models.
    (c) Reconstructed LDA projection vectors for 2, 10, and 30 PCA components, reshaped into image space to highlight spatial–spectral features most relevant for discrimination. 
    (d) Distributions of inner-product scores between the projection vectors and normalized DF images for both particle types, with optimal decision thresholds indicated. 
    Accuracies of 75\%, 91\%, and 95\% were achieved for 2, 10, and 30 PCA components, respectively.}
 
  \label{fig:LDA}
\end{figure}

{We first explored a PCA–LDA (Linear Discriminant Analysis) pipeline as a base model for distinguishing nanocubes from nanospheres solely using DF RGB images. As shown in Figs.~\ref{fig:LDA}(a) and (b), the classification accuracy improved with increasing PCA dimensionality, with a substantial rise between 10 and 30 components. Based on this, we evaluated three representative cases: 2, 10, and 30 PCA components, achieving accuracies of 75\%, 91\%, and 95\%, respectively.}

{To gain physical insight into the classification process, we reconstructed the LDA discriminant vectors into image space. These projection vectors, shown in Fig.~\ref{fig:LDA}(c), highlight spatial and spectral regions most relevant for discrimination. For example, in the 30D PCA case of Fig.~\ref{fig:LDA}(c), the green channel is stronger at the image center, while the red channel is weaker. Combined with the kernel density estimation (KDE) plots in Fig. \ref{fig:LDA}(d) where spheres have higher projection scores than cubes. This indicates that, statistically, nanospheres exhibit higher green intensity at the center of their DF images, whereas nanocubes display more pronounced red in the same region. Physically, when the overall dimensions are similar, this difference may arise from the distinct far-field distributions of the two geometries: nanospheres tend to support more symmetric scattering modes that enhance shorter-wavelength (green) scattering at the center, while nanocubes, with sharper edges and corners, can localize fields in a way that favors longer-wavelength (red) scattering in the central region \cite{chikkaraddy2017ultranarrow}. These projection vectors can also be used directly for classification by correlating them with normalized DF images and applying a single decision threshold (Fig. \ref{fig:LDA}(d)), effectively combining feature extraction and classification in one interpretable step.}

{For comparison, we trained a \textit{SPARX classifier}, whose architecture is shown in Fig.~S9. As illustrated in Fig.~\ref{fig:LDA} (a) with a red bar, this model achieved a classification accuracy of 99.8\%, substantially higher than the best PCA–LDA results. One likely reason for this improvement is that \textit{SPARX classifier}, as a convolutional neural network, is inherently more robust to variations in the spatial position of features within an image. In our case, this means that a DF image of a nanoparticle can be accurately classified even if its scattering pattern appears at a slightly different location in the frame. In contrast, methods such as LDA are sensitive to absolute pixel positions, and the same DF pattern shifted spatially within the image may be classified incorrectly.}

{In summary, we have demonstrated the \textit{SPARX} framework, which reveals the rich information hidden in simple DF RGB images. From these RGB signals, \textit{SPARX} can reconstruct the broadband spectra and, importantly, extract and understand information for specialized applications such as nanoparticle screening and shape classification. 
It is reasonable to anticipate that other properties related to DF scattering, such as particle size or nanogap thickness, could also be inferred directly from DF RGB images in the future. 
In contrast, conventional spectral acquisition methods, including HSI, can precisely capture spectra, yet they do not decode the underlying multimodal resonance correlations inherent to optical systems, limiting their suitability for inference tasks such as classification.}

However, our DL-based \textit{SPARX} reduces both cost and technical barriers: after training, it relies solely on standard dark-field microscopy components (e.g., an RGB camera) and does not require a spectrometer, whereas conventional hyperspectral systems typically demand more expensive, specialized hardware and stricter spectral–spatial calibration. 
{While variations in microscope optics, illumination, and camera response can affect measured RGB values, a limitation relevant for generalizability and commercial deployment, this issue can be largely mitigated in controlled laboratory setups using standardized optical components, which are common in commercial systems. Additional strategies such as transfer learning with small calibration datasets or domain-adaptive normalization could further enhance cross-hardware robustness. Furthermore, \textit{SPARX }leverages spatial intensity distributions and inter-channel correlations, which are inherently more resilient to hardware variations than absolute RGB values alone.}
This positions \textit{SPARX} as an accessible, lab-ready solution for general optical characterization and specialized plasmonic applications—validated across diverse nanoplasmonic geometries, including nanospheres and nanocubes as demonstrated in this work. And \textit{SPARX} is compatible with emerging techniques such as compressed sensing, enabling further acceleration when dealing with sparse samples.

\section*{Conclusion}\label{sec13}

We developed the \textit{SPARX} {framework}, a deep learning-based paradigm to rapidly and accurately extrapolate spectra from information-limited RGB DF images, {classify the shape of the nanoparticles,} beyond the RGB camera's capture capabilities. \textit{SPARX} batch-predicts nanoparticle spectra with millisecond latency, achieving a throughput {2-4} orders of magnitude faster than traditional serial spectroscopic measurements, while maintaining accuracy comparable to direct acquisitions. Beyond {spectral} prediction, \textit{SPARX} quantifies uncertainty using a heteroscedastic model, providing an additional key for the reliable screening of \textit{high-confidence} predictions. {Furthermore, the \textit{SPARX classifier} enables inference tasks for additional screening of the samples, which are challenging for conventional spectroscopic methods.} 

Therefore, \textit{SPARX} opens exciting possibilities for addressing a critical challenge in extreme nanophotonics: the intrinsic trade-off between exceptional single-particle performance and lack of reproducibility across a population, a long-standing bottleneck. By combining accurate extrapolation, batch processing, reliable uncertainty quantification, inference capability, and transplantability across diverse nanogap geometries, \textit{SPARX} paves the way for multi-task, high-throughput, and reproducible measurements. Beyond plasmonic applications, the general \textit{SPARX} paradigm---replacing spectrometer-dependent workflows with camera-based deep learning---establishes a scalable and lab-ready methodology for next-generation optical characterization and technology.

\section*{Methods}\label{methods}

\subsection*{Optical Setup}
\textbf{Automated DF microscopy and spectroscopic characterization}

\noindent The optical setup used to obtain the data for this study is illustrated in Figure \ref{fig:setup}(a). The DF imaging of NPoM systems was implemented using a commercial illuminator (Olympus BX53). A halogen lamp served as the white light source, collimated through a condenser lens and subsequently directed to a DF module containing a 45°-tilted annular mirror. This optical configuration generated annular illumination that was coupled into the outer annular channel of a DF objective (NA=0.9, working distance=1 mm), producing grazing-incidence excitation at the sample plane. Scattered light from individual NPoM nanostructures was collected through the same objective and subsequently split by a 50/50 beam splitter (Chroma). The reflected optical path was directed to an imaging module where a tube lens focused the signal onto an RGB CMOS camera (Tucsen MIchrome 20), enabling wide-field DF imaging. The transmitted path was coupled into a spectrometer (Horiba IHR320, {CCD: Symphony II FIVS}) through a motorized entrance slit, with spectral dispersion achieved via a grating (150 l/mm) and detection using a CCD. Spatially resolved single-particle spectroscopy was achieved through confocal alignment optimization, where the spectrometer slit width was precisely adjusted to match the dimension of individual nanoparticles. DF spectra were normalized using the expression $S = (A - B)/L$, where A represents the raw scattering spectrum, $B$ denotes the background spectrum acquired from adjacent mirror regions, and $L$ corresponds to the illumination source spectrum. An home-built automated measurement protocol was implemented for high-throughput characterization. Individual NPoM structures were localized via particle tracking algorithms and precisely positioned at the optical axis center using a motorized XYZ translation stage. Synchronized acquisition of both spectral and imaging data streams was achieved through custom control software, enabling correlated structural and optical analysis of single NPoM nanoantennas.
\subsection*{DF Data Processing}
The DF images are captured at a resolution of 128 × 128 pixels, while the spectral data is acquired within a wavelength range of 468 nm to 1026 nm, sampled at 2048 discrete points. This spectral data is down sampled to 128 points on  the spectral axis and intensity got normalized to the mean of the entire spectral data.

\subsection*{Supervised Learning}

{We develop a hybrid 2D--1D deep neural network to predict one-dimensional spectral responses from two-dimensional dark-field scattering images. The model takes $128 \times 128 \times 3$ images as input and outputs a spectral sequence of length 128, along with associated uncertainty estimates.The network follows an encoder--decoder architecture. The encoder consists of six hierarchical 2D convolutional stages, each comprising a residual convolutional block and a max-pooling layer (stride $2 \times 2$) for progressive multi-scale feature extraction. Each residual block is built from stacked $3 \times 3$ convolutions with Batch Normalization and ReLU activation, and incorporates L1--L2 regularization ($l_1 = 10^{-5},\; l_2 = 10^{-4}$) to enhance generalization.At the end of the encoder, high-dimensional features are flattened and projected through a fully connected layer, then reshaped into a one-dimensional latent representation, enabling the transformation from spatial to spectral domain.}

{The decoder comprises five 1D convolutional stages. Each stage applies a residual Conv1D block (kernel size 3) followed by transposed convolution (stride 2) for progressive upsampling, recovering the output sequence to length 128. A final Conv1D layer produces a two-channel output corresponding to the predicted spectrum and its uncertainty.}

{The dataset was divided into 8,382 images for training, 2,794 images for validation during training, and an independent test set of 1,000 images with paired spectral measurements used exclusively for final evaluation. The model is trained using the Adam optimizer with an initial learning rate of $1 \times 10^{-4}$, a batch size of 32, and 150 epochs. An adaptive learning rate scheduling strategy based on validation performance is employed to facilitate convergence.}

{To mitigate over-fitting, model training employed validation-based checkpointing and early stopping, with the best-performing model selected according to the minimum validation loss. Consistent training and validation losses and stable performance on an independent test set indicate good generalization.}

{For the classification study, nanoparticle labels were assigned according to the fabrication method and the shape and size distributions provided by the commercial supplier. Minor polydispersity or irregular by-products may be present but were not considered in this classification study. }

{Please find the source code, trained models, and the curated dataset available from the repository stated in the Data availability section.}
\subsection*{Multivariate Analysis}

To analyze our data, we employ PCA, a statistical technique that reduces high-dimensional data into a lower-dimensional space by identifying new orthogonal axes, principal components, that capture the maximum variance within the dataset. This transformation helps reveal patterns and correlations that may not be obvious in the original feature space. Since DF images are in RGB format, the PCA components of the DF images are normalized between 0 and 1 for better color visualization. To assess the reliability and structure of the DF image PCA projections, we applied several color mappings (see Figure \ref{fig:setup}):

\begin{enumerate}
    \item \textbf{Spectral PC1:} DF PCA points were colored by the first principal component of the spectral data. This highlighted strong alignment between image-based clusters and spectral variation.
    
    \item \textbf{Acquisition index:} Points were colored based on their sample indices, which correspond to the order of measurement over several days. The intermixing of colors across clusters indicates high reproducibility and minimal acquisition bias.
    
    \item \textbf{Spectral features:} We also used two key spectral metrics for coloring (i) the resonance wavelength ($\lambda$), and (ii) the total spectral energy. Both of which show that clusters in the DF projection map correspond to physically meaningful spectral properties.
\end{enumerate}

\subsection*{Heteroscedastic Loss Function for Spectral Modeling}

To explicitly model the heteroscedasticity in prediction error, we assume that at each wavelength $\lambda$, the predicted spectral intensity follows a Gaussian distribution centered at the predicted mean $\mu(\lambda)$ with a variance $\sigma^2(\lambda)$:

\begin{equation}
P(y(\lambda) | \mu(\lambda), \sigma^2(\lambda)) = \frac{1}{\sqrt{2\pi \sigma^2(\lambda)}} \exp\left( -\frac{(y(\lambda) - \mu(\lambda))^2}{2\sigma^2(\lambda)} \right),
\end{equation}

where $y(\lambda)$ is the ground truth intensity, $\mu(\lambda)$ is the predicted mean, and $\sigma^2(\lambda)$ is the predicted variance. The model is trained by minimizing the negative log-likelihood (NLL) of this Gaussian distribution, which leads to the following loss function:

\begin{equation}
\mathcal{L}_{\text{NLL}} = \frac{1}{N} \sum_{i=1}^{N} \left[ \log \sigma^2(\lambda_i) + \frac{(y(\lambda_i) - \mu(\lambda_i))^2}{\sigma^2(\lambda_i)} \right] + \text{const.}
\end{equation}

The constant term does not affect optimization and is thus ignored. To implement this, the final convolutional layer of the network is modified to output two values per wavelength: the mean $\mu(\lambda)$ and the log variance $\log \sigma^2(\lambda)$. This dual-output architecture allows the model to learn both the expected value and the uncertainty of the spectrum simultaneously.

\subsection*{Numerical Simulation}
Electromagnetic simulations were done with the commercial finite-element method package COMSOL Multiphysics. Since the nanoparticle on mirror system has an axis-symmetry, we implemented so-called 2.5D calculation method \cite{ciraci2012probing} on that with an oblique incident light (considering 0.9 NA). Permittivity of the gold follows the famous Olmon et.al, dataset \cite{olmon2012optical}. The scattering spectra of the nanoparticles were calculated by integrating the energy flow of the scattered field. We considered that the molecule layer has a refractive index of 1.4 situated between the nanoparticle and the film to acting as insulator to form a metal-insulator-metal nanocavity.

\subsection*{PCA–LDA Classification Pipeline}
{For PCA–LDA classification, DF RGB images (originally 128×128 pixels) were resized to 32×32 pixels to reduce computational cost. Each image was then flattened from (32, 32, 3) to a vector of length 3072 before PCA transformation. PCA reduced the data to $n$ components, which were then passed to an LDA classifier. The number of PCA components was varied to study the effect on classification accuracy.}

{For discriminant visualization, the LDA projection vector was mapped back to the original data space by multiplying it with the PCA transformation matrix and then reshaping it to (32, 32, 3). These projection vectors were used both for visual interpretation and as direct classifiers. For direct classification, the inner product between the projection vector and a normalized DF image was computed, and a threshold was applied to determine the particle type. Score distributions for nanospheres and nanocubes, along with optimal thresholds, are provided in Fig.~\ref{fig:LDA}.}

\subsection*{Sample preparations}
Gold mirror was fabricated using the template-stripped method \cite{nagpalUltrasmoothPatternedMetals2009}. Briefly, 100-nm-thick Au films were thermally evaporated onto a clean silicon wafers ($\sim{0.5}\;\rm{nm\cdot s^{-1}}$ deposition rate) followed by epoxy bonding to quartz substrates using UV-curable optical adhesive (NOA61, Norland Products). Mechanical cleavage at the silicon-quartz interface using a precision razor blade exposed atomically smooth Au surfaces, with root-mean-square roughness about 0.3 nm \cite{chenProbingSubpicometerVertical2018}.
CTAC stabilized Au nanoparticles (80 nm diameter, $\sim{0.1}\;\rm{mg\cdot mL^{-1}}$ aqueous dispersion) were purchased from Micetech Co. Ltd. For NPoM assembly, $\sim 1\;\mu\rm{L}$ nanoparticle solution were drop-casted onto freshly stripped Au mirrors and incubated for 5 min under ambient conditions. Substrates were subsequently dried under nitrogen flow and immersed in ultrapure water with 10 s to remove excess CTAC on the sample surface, followed by secondary nitrogen drying. This protocol yielded NPoM structures with self-assembled $\simeq$ 1–2 nm CTAC spacer layers \cite{chenProbingSubpicometerVertical2018}, forming well-defined plasmonic nanogap structures.

\section*{ASSOCIATED CONTENT}
\subsection*{Notes}
The authors declare no competing interests. {Views and opinions expressed are however those of the author(s) only and do not necessarily reflect those of the European Union or the European Research Council. Neither the European Union nor the granting authority can be held responsible for them.}
{A preprint version of this manuscript is available.\cite{kazemzadeh2025}}

\subsection*{Authors' contributions}
M. Kazemzadeh and B. Zhang contributed equally to this work.
H. Hu conceived the idea and did electromagnetic simulations. M. Kazemzadeh trained the deep learning model and analyzed the data. B. Zhang, W. Jiang, and W. Chen performed the experimental investigation and collected data for the training. W. Chen, B. Zhang, T. He, H. Liu, Z. Jiang, Z. Hu, X. Dong, C. Sun, X. He, and H. Xu contributed to the establishment of the automatic collection optical setup. 
M. Kazemzadeh and H. Hu created the figures and wrote the original draft, reviewed and edited. All the authors discussed, reviewed, and edited the paper.

\subsection*{Data availability}
{All the source code, trained models, and the curated dataset are available at {{https://doi. org/10.5281/zenodo.19499234}}.} All other data that support the plots within this paper and other findings of this study are available from the corresponding author upon reasonable request.



\subsection*{{Funding Sources}}

{This work was supported by the National Key Research and Development Program of China (Grant No. 2024YFA1409900) and the National Natural Science Foundation of China (Grant No. 62475071 and 52488301). M.K., and F.P. acknowledge funding by European Union's Horizon Europe under grant agreement 101125498, project MINING - “Multifunctional nano-bio interfaces with deep brain regions”. M.K., and F. P. acknowledge funding from the European Union's Horizon 2020 Research and Innovation Programme project DEEPER under grant agreement 101016787.}

\begin{suppinfo}

Please find a detailed explanation of the comparison of DF images and spectra of DL selections, UMAP unsupervised learning, \textit{SPARX} architecture, homoscedastic \textit{SPARX} model performance evaluation, generality test of \textit{SPARX} on different plasmonic systems, and unsupervised learning on the nanoparticle shapes in the Supporting Information.

\end{suppinfo}

\bibliography{achemso-demo}
\newpage
{\section*{For Table of Contents Use Only}}
\noindent \textbf{Seeing Beyond RGB Capabilities: Data-Driven and Physics-Guided Broadband Spectral Extrapolation of Plasmonic Nanostructures by Deep Learning}

\noindent Mohammadrahim Kazemzadeh, Banghuan Zhang, Tao He, Haoran Liu, Zihe Jiang, Zhiwei Hu, Xiaohui Dong, Chaowei Sun, Wei Jiang, Xiaobo He, Shuyan Li, Gonzalo \'Alvarez-P\'erez, Ferruccio Pisanello, Huatian Hu, Wen Chen, and Hongxing Xu

\subsection*{Description:}
\noindent \textit{SPARX} deep learning model can reconstruct and extrapolate spectra of the plasmonic nanostructures from their information-limited RGB images.

\begin{figure}[h]
\centering\includegraphics[width=8.25cm]{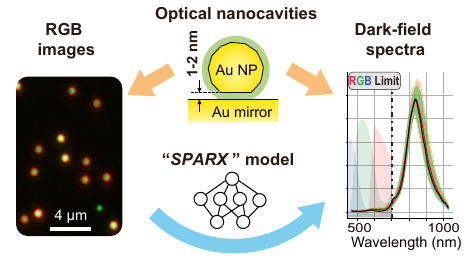}
\end{figure}
\end{document}